\documentstyle[aps,prd,epsfig]{revtex}
\def\lsim{\raise0.3ex\hbox{$<$\kern-0.75em\raise-1.1ex\hbox{$\sim$}}}
\def\gsim{\raise0.3ex\hbox{$>$\kern-0.75em\raise-1.1ex\hbox{$\sim$}}}
\makeatletter
\@addtoreset{equation}{section}
\makeatother

\newcommand{\beqn} {\begin{equation}}
\newcommand{\eqn} {\end{equation}}
\newcommand{\bc} {\begin{center}}
\newcommand{\ec} {\end{center}}

\newcommand{\slsh}[1] {#1\kern-.43em/}
\newcommand{\real}{{\sf I}\kern-.12em{\sf R}}
\newcommand{\comp}{{\sf I}\kern-.48em{\sf C}}
\newcommand{\nin} {\in\kern-.6em/}

\newcommand{\al}{\alpha}
\newcommand{\be}{\beta}
\newcommand{\ga}{\gamma}
\newcommand{\de}{\delta}
\newcommand{\ep}{\epsilon}

\def\MEF{m_{\rm eff}}\def\mef{\ifmmode\MEF\else$\MEF$\fi}

\def\SM{s_{\mu}}\def\xm{\ifmmode\SM\else$\SM$\fi}

\begin{document}
\thispagestyle{empty}
\vskip -100pt
\mbox{} \hfill BI-TP 98/09\\[-3mm]
\mbox{} \hfill August 1998\\[-3mm]
\mbox{} \hfill revised version
\begin{center}
{{\large \bf Diquark Masses from Lattice QCD}
\medskip
} \\
\vspace*{0.5cm}
{\large M. He{\ss}, F. Karsch, E. Laermann and I. Wetzorke} \\
%
{\normalsize
$\mbox{}$ {Fakult\"at f\"ur Physik, Universit\"at Bielefeld,
D-33615 Bielefeld, Germany}
}
\end{center}
\vspace*{1.0cm}
\centerline{\large ABSTRACT}

We present first results for diquark correlation functions calculated in
Landau gauge on the lattice. Masses have been extracted from the long distance
behaviour of these correlation functions. We find that the ordering of  
diquark masses with spin 0 and 1 states in colour anti-triplet and 
sextet channels is in accordance with instanton motivated interaction 
models. 
Although we find evidence for 
an attractive interaction 
in 
colour 
anti-triplet states with a splitting between spin 0 and spin 1  
diquarks that can account for the mass splitting between the 
nucleon and the delta, there is
no evidence for a deeply bound 
diquark state.

\section{Introduction}

Since a long time it has been speculated that QCD at finite baryon number
density and low temperature may have a much richer phase structure than at
high temperature and low or vanishing baryon number density. In general
this is due to the fact that quantum statistics is much more important
at low temperature and energetically favours bosonic forms of matter
over fermionic matter. At low temperature and low density the
possibility of pion and kaon condensates thus has been discussed. Unlike
in the high temperature case, where it is evident that the strongly
interacting matter exists in the form of a quark-gluon plasma it has
been speculated that at high baryon density bosonic states -- diquarks and
even larger quark clusters -- may play an important role \cite{Don88}.
Recently it has been suggested that diquarks may, in fact, form a Bose
condensate
at high densities and low temperatures \cite{Rap97,Alf98} and also the
possibility of a dibaryon phase has again been discussed \cite{Fae97}.

The realization of such states of matter crucially depends on details of the
interaction among quarks. In particular, the analysis of the fine structure
of the hadron spectrum such as the nucleon-delta mass splitting suggests
that there exists a strong attractive force between quarks in a colour
anti-triplet state with anti-parallel spin orientation which leads to the
formation of a diquark state. Such a spin dependent interaction naturally
arises in the framework of perturbative QCD from one-gluon exchange
\cite{deR75}.
Furthermore it has been realized that a spin and flavour dependent interaction
among constituent quarks is needed to account for a satisfactory description of
the fine structure of the experimentally observed baryon spectrum \cite{Glo96}.
Such an attractive spin and flavour dependent
interaction indeed is induced by instantons \cite{tHo76,Shu89}.
This raises the possibility for the existence of rather tightly bound spin 0
diquark states. Some indications for such states have been found
from the analysis of diquark correlation functions in the instanton
liquid model \cite{Sch94}.
A quantitative analysis of the interaction between quarks is clearly needed to
judge the existence of the interesting diquark phase structure discussed in
\cite{Rap97,Alf98}. In particular, the possible coexistence of a chiral
symmetry broken phase with a diquark condensate will crucially depend on the
value of the diquark masses.

In this letter we will present first results from a calculation of diquark
correlation functions on the lattice. In particular we present results
for diquarks in the colour anti-triplet representation.
Our spectrum calculations have been
performed in quenched QCD using perturbatively improved gauge and fermion
actions. As quark and diquark correlation functions are gauge-variant
observables we have performed all our calculations in a fixed
gauge\footnote{Alternatively one could couple the light
diquark states to a heavy quark, which serves to neutralize the colour
\cite{Sch94}. This, however, only trades {\it gauge dependence} against
{\it path dependence}.}.
We have used the Landau gauge as is commonly done in the
analysis of quark \cite{Ber90,Sku95} and gluon masses \cite{Nakxx,Hel95}
on the lattice.

\section{Quark and Diquark Correlation Functions}

While it is generally expected that the interaction between up and down
quarks in a spin 0 colour anti-triplet state is attractive, it is not
that obvious how quarks in other quantum number channels interact.
The analysis of diquark states in the instanton liquid model suggests
a quite deeply bound spin 0 diquark state \cite{Sch94}. Such calculations,
however, may miss a contribution from the confining part of the quark-quark
interaction. In general the diquark masses will receive contributions
from the constituent masses, the confining and the spin-dependent part
of the quark-quark interaction. In terms of a simple potential model
these contributions are to leading order additive \cite{Glo96},
\begin{equation}
m_{FSC} = 2 \bar{m}_q + V_{conf} + V_{S}\;,
\label{potential}
\end{equation}
where the subscript $FSC$ denotes {\it flavour, spin and colour} quantum
numbers, $\bar{m}_q$ denotes the constituent quark mass, $V_{conf}$ and  $V_{S}$
are the contributions from the confining and spin-dependent parts of the
quark-quark potential. The latter may include contributions from
flavour-spin as well as colour-spin dependent interaction terms.
Some information on the contribution of the different terms to the diquark
masses exists from the analysis of hadron masses within the framework
of potential models. In the case of three quark states the corresponding
form of Eq.~\ref{potential} reads, for instance,
$m_N= 3 \bar{m}_q + \tilde{V}_{conf} + 2V_0 + V_1$
for the nucleon and $m_{\Delta} = 3 \bar{m}_q + \tilde{V}_{conf} + 3V_1$ for
the delta, with $\tilde{V}_{conf}= 3 V_{conf}$.
While the masses are expected to receive the largest contributions from
the constituent mass term ($\bar{m}_q \simeq 300$~MeV) and the confining part
($V_{conf} \simeq 200$~MeV \cite{Glo96}) the nucleon-delta mass splitting is
entirely determined by the difference in the spin dependent part of the
potential. The latter also is related to the mass splitting of the $S=0$ and $S=1$
diquarks,
\begin{equation}
m_{613} - m_{303} = {1\over 2} \bigl(m_\Delta -m_N \bigr)\;.
\label{mdif}
\end{equation}

In the following we will analyze $S=0$ and $S=1$ diquark states in different
colour and flavour representations. The four different states considered
are listed in Table~\ref{tab:states}.

The corresponding diquark correlation functions are built up
from the quark correlation function $G^{ab}_{\alpha,\beta}$. For the
colour anti-triplets we obtain, for instance,

\begin{table}
\begin{center}
\begin{tabular}{cccc}
$(F,S,C)$&state&F-S coupling&C-S coupling  \\
\hline
$(3^*,0,3^*)$&
$\ep_{abc} (C\ga_5)_{\al\be}u_{a,\al}^{\dagger}d_{b,\be}^{\dagger}$
&-2&-2 \\
$(6,1,3^*)$&$\ep_{abc}u_{a,\alpha}^{\dagger}u_{b,\alpha}^{\dagger}$
&-1/3&2/3 \\
$(3^*,1,6)$&$u_{c,\alpha}^{\dagger}d_{c,\alpha}^{\dagger}$
&2/3&-1/3 \\
$(6,0,6)$&$(C\ga_5)_{\al\be}u_{c,\al}^{\dagger}u_{c,\be}^{\dagger}$
&1&1 \\
\end{tabular}
\end{center}
\caption{Diquark states with spin $S$ in flavour ($F$) and colour
($C$) anti-triplet and sextet representations. The third and fourth
column give the relative strength of interaction terms corresponding to
a flavour-spin and colour-spin coupling, $V_S \sim (\lambda_1^a \lambda_2^a)
(s_1 s_2)$, where $\lambda_i^a$ denote the generators of $SU(3)_{flavour}$
or $SU(3)_{colour}$, respectively.}
\label{tab:states}
\end{table}

\begin{eqnarray}
G^{303}_{cf} (\vec{x},t) &=&
 \ep_{abc} \ep_{def} (C\ga_5)_{\al\be} (C\ga_5)_{\ga\de}
         G_{\al \ga}^{a d} G_{\be \de}^{b e} \nonumber \\
G^{613}_{cf} (\vec{x},t) &=&
 \ep_{abc} \ep_{def} (G_{\alpha \beta}^{a d} G_{\alpha \beta}^{b e}
         - G_{\alpha \beta}^{a e} G_{\alpha \beta}^{b d})
\label{correlator}
\end{eqnarray}
where Latin (Greek) indices denote colour (spinor) degrees of freedom.

We have analyzed these correlation functions in Landau gauge. To be
specific, we have calculated the diagonal correlators,
$G^{F,S,C} (t)\equiv  G^{F,S,C}_{a,a}$ and the
scalar part of the quark propagator,
$G_q (t) \equiv  4 G^{a,a}_{\alpha,\alpha}$ where the sum is taken
over $a$ and $\alpha$ and also over the spatial coordinates $\vec{x}$
in order to project onto zero-momentum states.

\begin{figure}[htb]
\bc
\epsfig{file=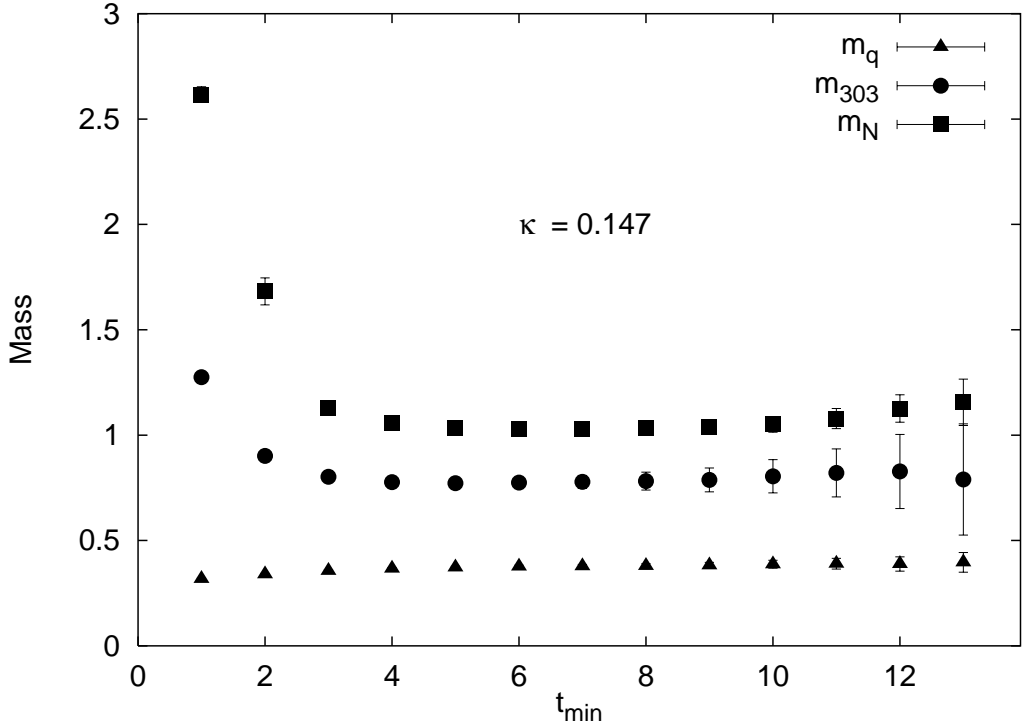,height=10cm}
\ec
\caption{Masses extracted from single exponential fits to the
correlation functions of quark, colour anti-triplet spin 0 diquark and
nucleon at $\kappa=0.147$ in the interval $[t_{\rm min},32-t_{\rm min}]$.
}
\label{fig:mass}
\end{figure}

Our calculations have been performed on lattices of size $16^3\times 32$.
The gauge field configurations have been generated with a tree-level
Symanzik improved action at a gauge coupling $6/g^2= 4.1$. A calculation
of the string tension at this value of the coupling leads to
$\sqrt{\sigma}a=0.3773(22)$, i.e. a cut-off $a^{-1} \simeq 1.1 $~GeV
\cite{Bei97}\footnote{Here and in the following we use
$\sqrt{\sigma}=420$~MeV to set the scale for all masses. Although
$\sqrt{\sigma}$ is experimentally less well determined than a hadron
mass it is more appropriate to use it to set the scale in a
quenched calculation where hadron masses are calculated for various
values of the quark mass}. \pagebreak We have generated 73 gauge field
separated by 100 sweeps of 4 overrelaxation and 1 heat bath steps each. These
are fixed to Landau gauge using an algorithm based on Fourier accelerated
overrelaxation \cite{Ranxx}. 
In the fermion sector we use the Sheikholeslami-Wohlert action with a
tree-level clover coefficient \cite{Sheik}. On each gauge field configuration
the fermion matrix has been inverted four times, {\it i.e.} with source vectors
at four different lattice sites, for eight different quark mass values.
Our analysis thus is based on a sample of 292 quark propagators.
Fits have been performed with one and two exponential functions.
Leaving out successively data points at small time separations we look for
stable results for the fitted masses.
Typical results obtained from single exponential fits of quark, diquark and
nucleon correlation functions are shown in Figure~\ref{fig:mass}. Fits
with two exponential functions do reach a plateau earlier and yield
consistent results.

\pagebreak

In addition to the gauge dependent quark and diquark correlation functions
we also construct the standard hadron correlation functions for the
pion, rho, nucleon and delta. The 8 different quark masses selected
correspond to the interval $0.5 < m_{\pi}/m_{\rho} < 0.9$.
Our current
analysis thus is still restricted to rather heavy quark masses. This is also
reflected in the fact that the ratio $m_N / m_{\rho}$ is still close
to that of the additive quark model, {\it i.e.}  $m_N / m_{\rho} \simeq 1.5$.
Nonetheless we note that our analysis does yield quite sizeable
results for the nucleon-delta mass splitting (Table~\ref{tab:masses})
which was determined from a simultaneous fit of the ratio of
nucleon and delta correlation functions and the nucleon correlation functions.
From an extrapolation to the chiral limit we obtain
$m_\Delta - m_N = 184~(63)$~MeV, which is about 40\% below the experimental value.
It is, however, consistent with earlier calculations with similar
parameters \cite{DeGra91,Wein94}.
For our current analysis it is reassuring that we can
observe a sizeable splitting in baryonic states and are thus
sensitive to the spin dependence of hadron masses.
\\\\
\begin{table}
\begin{center}
\begin{tabular}{lllll}
$\kappa$&$m_\pi$&$m_\rho$&$m_N$&$m_\Delta - m_N$  \\
\hline
0.140 & 0.910(1) & 1.025(2)  & 1.608(6)  & 0.065(10) \\
0.142 & 0.794(1) & 0.932(2)  & 1.457(6)  & 0.079(12) \\
0.144 & 0.667(1) & 0.836(3)  & 1.298(6)  & 0.098(16) \\
0.145 & 0.596(1) & 0.787(4)  & 1.212(7)  & 0.114(20) \\
0.146 & 0.519(1) & 0.739(7)  & 1.122(10) & 0.128(22) \\
0.147 & 0.430(2) & 0.688(15) & 1.029(13) & 0.135(41) \\
0.1475& 0.379(2) & 0.661(18) & 0.982(13) & 0.131(66) \\
0.148 & 0.316(3) & 0.595(66) & 0.924(14) & 0.147(97) \\
\hline
$\kappa_c$& ~    & 0.579(18) & 0.821(13) & 0.166(56) \\
\end{tabular}
\end{center}
\caption{Hadron masses and the nucleon-delta mass splitting for
various values of the hopping parameter. The last row gives
the results of an extrapolation to the chiral limit obtained from
a linear fit in $\kappa^{-1} - \kappa_c^{-1}$. The critical value of the
hopping parameter for vanishing pion mass has been determined as
$\kappa_c = 0.14923(2)$.}
\label{tab:masses}
\end{table}

The basic component for the analysis of diquark correlation functions
is the quark propagator. Although the quark mass is, in principle, a
gauge-variant quantity, it has been found to show only little
gauge dependence in a class of covariant gauges, which includes the
Landau gauge \cite{Ber90}. The quark masses extracted in Landau gauge
are consistent with constituent quark mass values of (350-400)~MeV
\cite{Ber90,Sku95}. Our results for the quark correlation function
are consistent with these earlier findings. We generally observe that
local masses extracted from $G_q (t)$ rise with increasing temporal
distance $t$ and develop a plateau for $t\gsim 6$. In this region we
have performed exponential fits to extract the quark mass. The results
are summarized in the second column of Table~\ref{tab:quark}. Using
the five largest $\kappa$-values to extrapolate linearly in
$\kappa^{-1} - \kappa_c^{-1}$ to the chiral limit, we find
$\bar{m}_q / \sqrt{\sigma} = 0.813(30)$ or $\bar{m}_q \simeq 342(13)$~MeV.
\\\\
\begin{table}
\begin{center}
\begin{tabular}{llll}
$\kappa$&$\bar{m}_q$&$m_{303}$&$m_{613}$  \\
\hline
0.140 & 0.586(5)  & 1.190(9)  & 1.207(11) \\
0.142 & 0.531(5)  & 1.079(12) & 1.102(13) \\
0.144 & 0.472(6)  & 0.962(14) & 0.993(14) \\
0.145 & 0.442(6)  & 0.901(15) & 0.936(15) \\
0.146 & 0.410(8)  & 0.839(15) & 0.880(26) \\
0.147 & 0.378(11) & 0.774(15) & 0.827(33) \\
0.1475& 0.361(12) & 0.737(18) & 0.806(35) \\
0.148 & 0.344(11) & 0.696(18) & 0.806(45) \\
\hline
$\kappa_c$ & 0.307(11) & 0.623(19) & 0.727(37) \\
\end{tabular}
\end{center}
\caption{Quark and diquark masses for
various values of the hopping parameter.
The last row gives
the results of an extrapolation to the chiral limit obtained from
a linear fit in $\kappa^{-1} - \kappa_c^{-1}$. }
\label{tab:quark}
\end{table}

\begin{figure}[htb]
\bc
\epsfig{file=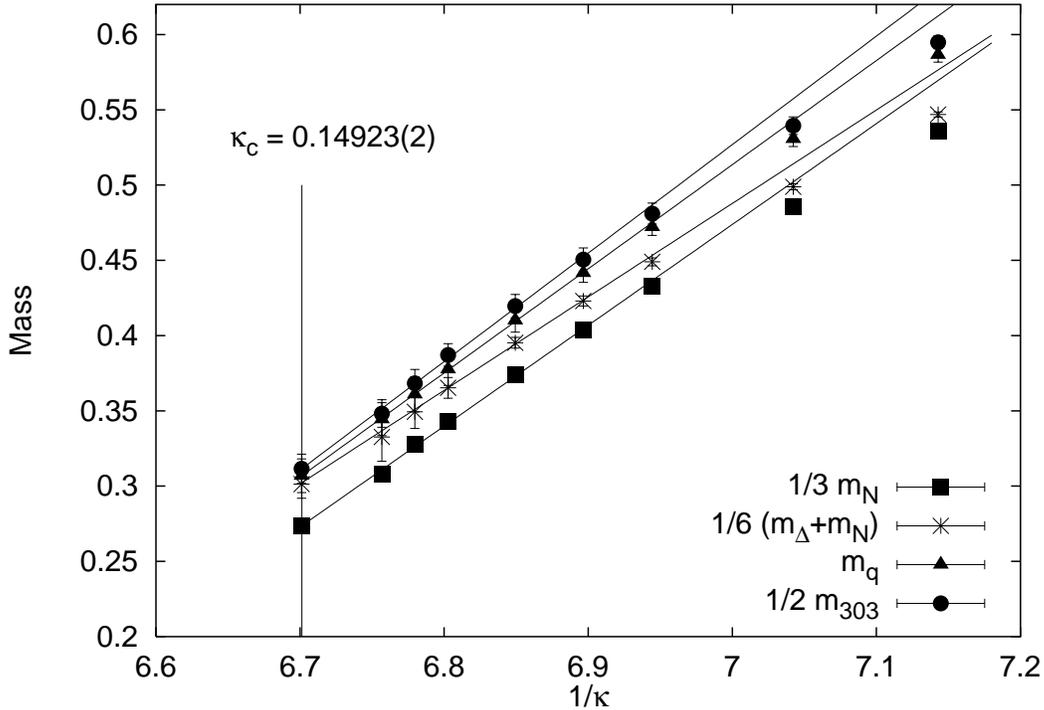,height=10cm}
\ec
\caption{1/2 diquark(303), 1/3 nucleon and 1/6 (nucleon+delta)
in comparison with the quark mass calculated in Landau gauge.}
\label{fig:bound}
\end{figure}

The diquark correlation functions corresponding to the colour anti-triplet
representation show a remarkably clean exponential decay. This is
reflected in the small $t$-dependence of local masses shown in
Figure~\ref{fig:mass} and also suggests the existence of an attractive
interaction in this channel. Although the diquark correlation functions are
also gauge 
variant the local masses show a behaviour very similar to that of ordinary
hadron masses. They approach a plateau from above. In the case of the spin 0
diquark state this is typically reached already for $t \gsim 4$. For the spin 1
correlation functions we observe larger contributions from excited states
at short distances and consequently the plateau is reached only for
$t \gsim 6$. The situation is improved when we perform fits with two
exponentials. These yield stable results for distances $t\gsim 2$.
The mass values obtained from such fits are given in Table~\ref{tab:masses} and
\ref{tab:quark}. In Figure~\ref{fig:bound} we show results for
half the spin 0 diquark masses and a combination of nucleon and delta masses,
$(m_\Delta + m_N)/6$, which generally is used as a definition of the
constituent quark mass. 
As expected from potential models $m_N/3$ is significantly lighter than 
the quark masses calculated by us in Landau gauge. We also note that
in the chiral limit ($\kappa \equiv \kappa_c$) the quark mass agrees
quite well with $(m_\Delta + m_N)/6$.
This is quite astonishing, as this   
phenomenologically defined constituent quark mass receives,
at least in the context
of potential models, additional contributions from the
confining part of the potentials as well as the spin-dependent parts.
This difference also is reflected in the different hopping parameter
(bare quark mass) dependence seen in Figure~2.

The mass of the spin 0, colour anti-triplet diquark is slightly larger
than twice the constituent quark mass.
Interpreted again in terms of a potential model this suggests that the 
positive energy contribution resulting from confinement is just balanced by
a contribution from an attractive spin interaction.
From Figure~\ref{fig:bound} as well as Table~\ref{tab:masses} and \ref{tab:quark}
it is, however, obvious that the diquark masses show no indication for a deeply bound
diquark state.
In the tables we also give the result of an extrapolation
to the chiral limit which is based on the five lightest quark masses and is
obtained from a fit linear in $\kappa^{-1} - \kappa_c^{-1}$.
Using again the string tension to set the scale we find
$m_{303}= 694~(22)$~MeV.

\begin{figure}[htb]
\bc
\epsfig{file=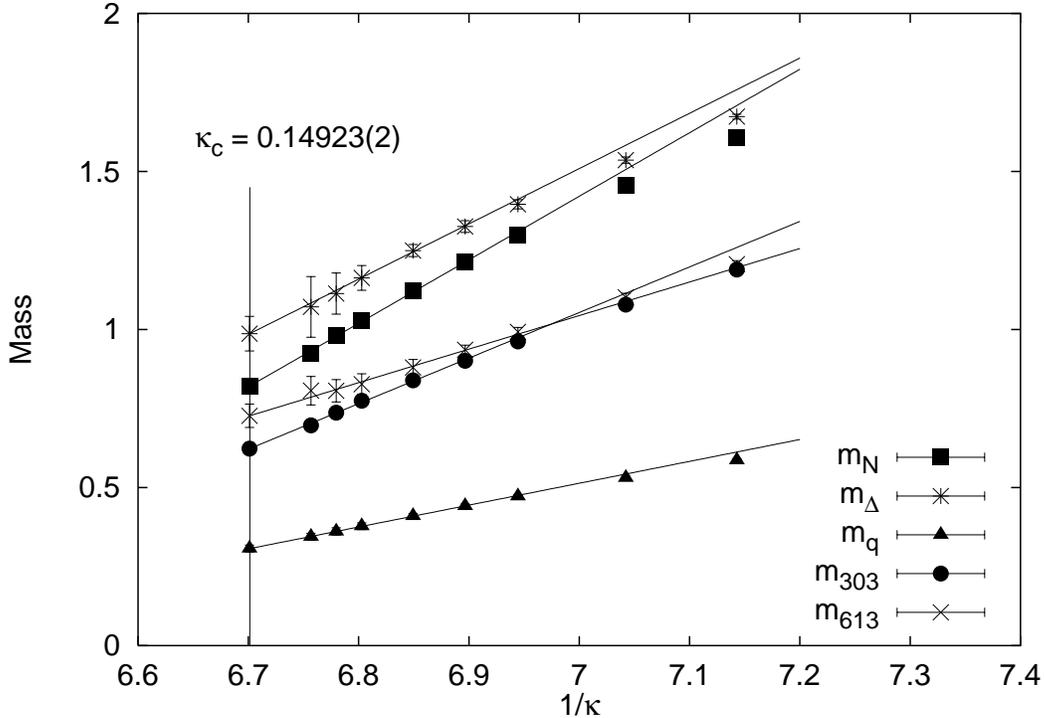,height=10cm}
\ec
\caption{Mass splitting between colour anti-triplet diquarks and nucleon/delta:
The lines show fits to the results obtained with the five lightest quark masses.}
\label{fig:split}
\end{figure}

Similar to what has been observed in
the case of the delta-nucleon mass splitting we find that the difference
between the masses of the spin 0 and spin 1 diquarks increases with
decreasing quark mass. This is shown in Figure~\ref{fig:split}.
From an extrapolation to the chiral limit we
find $m_{613} - m_{303} = 0.104(42)$, which in fact is about $60\%$ of the
splitting found in the nucleon channel.
This agrees well with the behaviour expected from potential models
(Eq.~\ref{mdif}).

%

Let us finally comment on the correlation functions for diquarks
in a colour sextet representation. We find that these have a much
smaller amplitude and are therefore much more noisy. Moreover, they receive
much larger contributions from excited states at short distances and reach
a plateau only very late.
The general tendency is that the colour-sextet states have larger masses than the
triplet states. This is in accordance with expectations based on a
flavour-spin dependent coupling (Table~\ref{tab:states}). We will report on a
more detailed analysis of these correlation functions elsewhere.

\section{Conclusion}
We have analyzed colour anti-triplet diquark states with spin 0 and 1 as well
as quark masses in Landau gauge. We find that the mass splitting between spin 0
and spin 1 diquark states can account for the observed mass splitting between
the nucleon and delta.

Although the current analysis has been performed in the quenched approximation
of QCD on quite coarse lattices and with fairly large quark masses
($m_{\pi}/m_{\rho}\gsim 0.5$) the current analysis does seem to rule out a
deeply bound diquark state. For the lightest spin 0 state we find
$m_{303}\simeq 700$ MeV which is slightly larger than twice the constituent
quark mass.

This makes the existence of a diquark phase coexisting with a chiral symmetry
broken phase as suggested in \cite{Rap97} unlikely. 
However, our finding of a sizeable splitting between the spin 0 and 1 anti-triplet diquark states
and the clear exponential decay of the corresponding correlation functions
gives some support for an 
attractive q-q interaction in the spin 0 channel and for
the existence of a colour superconducting diquark phase at high density.
We should stress that our current analysis has been performed in the $T=0$
confining phase of QCD where the attractive interaction due to instantons
(flavour-spin coupling) is expected to give the dominant contribution. In the high
density regime it is expected that the instanton induced interactions become
suppressed and one-gluon exchange (colour-spin coupling) becomes
increasingly important for the attractive interaction. Further studies of the
temperature and density dependence of diquark masses as well as the
nucleon-delta splitting are thus needed.
The latter will be difficult to analyze for 3-colour QCD due to the
well-known algorithmic problems in QCD with non-vanishing chemical
potential. The density dependence of the q-q interactions may, however,
first be analyzed in 2-colour QCD \cite{Rap97}.

\section*{Acknowledgements}

\vspace*{-0.2cm}

This work was partly supported by the TMR network {\it Finite Temperature
Phase Transitions in Particle Physics}, EU contract no. ERBFMRX-CT97-0122.
The work of F.K. was partly supported by a NATO Collaborative Research Grant,
contract no. 940451.


%
\end{document}